# Cryptovirology: Virus Approach

Shivale Saurabh Anandrao

Dept. of Computer Engineering,
Vishwakarma Institute Of Information Technology,
Pune-48, India
shivalesaurabh@yahoo.com

## Abstract

*Traditionally, "Cryptography" is a benediction to information processing and communications, it helps people to store information securely and the private communications over long distances. Cryptovirology is the study of applications of cryptography to build the malicious software. It is an investigation, how modern cryptographic tools and paradigms can be used to strengthen, develop and improve new malicious software attacks. Cryptovirology attacks have been categorized as : give malware enhanced privacy and be more robust against reverse-engineering, secondly give the attacker enhanced anonymity while communicating with deployed malware. This paper presents the idea of ``Cryptovirology'' which introduce a twist on how cryptography can also be used offensively. Being offensive means, it can be used to mount extortion based attacks that cause loss of access to information, loss of confidentiality, and information leakage, tasks which cryptography usually prevents. Also analyze threats and attacks that misuse of cryptography can cause when combined with fraudulent software (viruses, Trojans). Public-key cryptography is very essential for the attacks that based on cryptovirology. This paper also suggest some of the countermeasures, mechanisms to cope with and prevent such attacks. Even if the attackers actions on the host machine are being monitored, it still cannot be proven beyond reasonable doubt that he or she is the attacker; and it is an "originator-concealing attack". Evidence should be collected from the "author's own system which was used for the attack". These attacks have implications on how the use of cryptographic tools and techniques should be audited and managed in general purpose computing environments, and imply that access to the cryptographic tools should be in well control of the system(such as API routines). The experimental virus would demonstrate how cryptographic packages can be packed into a small space, which may have independent existence. These are many powerful attacks, where the attacker can encrypt the victim's data for ransom and release it after hostage.*

## *Keywords*:

*Malware, cryptovirology ,cryptovirus, cryptotrojan, electronic warfare threats, electronic espionage threats, evasive/ deniable/ untraceable attacks, password (information) snatching, public-key cryptography, kleptography, cryptanalysis,Gpcode.ak, Conficker, cryptoviral extortion.*

## 1. Introduction

Basically, Cryptovirology was born in academia [1]. However, practitioners have recently expanded the scope of this field to include the analysis of cryptographic algorithms used by malware writers, attacks on these algorithms using automated methods and analysis of viruses' encryptors. "Cryptovirology" is the study of the applications of cryptography to write malicious software[2][3]. Ciphers protect the system against the passive eavesdropper [3]. The Public key infrastructures protect against an active adversary that mounts a man-in-the-middle attack. Digital signature algorithms protect from a forger. E-cash systems protect from a counterfeiter and a 'double-spender' in E-transactions. Pseudorandom bit generators protect from a next-bit predictor. Cryptovirology can be extends beyond to finding protocol failures and design vulnerabilities [04][05]. It is forward engineering discipline that can be used for attacking purposes rather than defending. Understanding the possibility of the future attacks is the key to successfully protecting against them. Practitioners of protection mechanisms need to understand the potential ferocity and sophistication of viruses that are exist. Cryptovirology attacks have been aimed to give the malware privacy in greater extent and be more robust against getting caught also to give the attacker more anonymity while communicating with deployed malicious program, improve its capability to steal data from victims computer system, enhance the ability to carry out the extortion attack, enable new denial-of-service(DoS) attacks, enhance the fault-tolerance in distributed (network based) cryptoviral attacks. Also, recent work shows how a worm can install a backdoor on each infected system that opens only when the worm is presented with a system-specific ticket that is generated as a signal by the attacker. This is called as 'access-for-sale' worm. The current trends in computer attacks are increasing sophistication of attack tools and techniques, high speed of automation, vulnerability discovery rate that is hard to keep up with, increasing permeability of firewalls and highly asymmetric nature of threats new cryptoviruses. Cryptoviral extortion is a two-party protocol between the attacker and a victim that is carried using a cryptovirus, cryptoworm, or cryptotrojan[06].

This field was born with the observation that the public-key cryptography can be used to break the symmetry between what an antivirus analyst knows regarding a virus and what the virus writer. The former can only see a public key whereas the latter can see a public key and corresponding private key as well. The first attack that was identified in this field is the "cryptoviral extortion". The field includes hidden attacks in which the attacker secretly steals private information like private keys. In the cryptoviral extortion attack a malware hybrid encrypts the plaintext from the victim's system using the public key of the attacker. The attacker demands some form of payment from the victim as ransom in return for the plaintext that is held hostage. The set of attacks that presented, involves the unique use of strong cryptographic techniques with computer virus and the Trojan horse technology. They demonstrate how cryptography can be used by virus writer to gain explicit access control over the data that his virus has taken from the infected machines. Computer virus authors have made use of cryptography to make their creations more difficult to detect or analyze, or as part of the destructive payloads they carry. While cryptography does not play important role in most of the viruses presently responsible for the virus problem, but it is worth at least a passing glance [07][08][09].

Commonly encryption in computer viruses is as part of polymorphism. A polymorphic virus is one that changes its form as it spreads, so that two different files infected by the virus will not necessarily have any significant byte-strings in common. In machine-language viruses, polymorphism is generally achieved by splitting the virus into three different sections those are a cryptographic key, the decryptor code, and functional part of the virus. When the virus creates a new copy of itself, it selects a new cryptographic key, encrypts the main functional part of itself with that new key, and generates a new implementation of the decryptor code as a result. When the virus executes, the decryptor code runs and uses the key to decrypt the main functional part of the virus, which then takes control. Changing of the key changes the bytes in the encrypted body of the virus, and because key and the decryptor code vary with each copy of the virus, now there are two different instances of the virus will contain very different bytes-strings. This technique of encryption for polymorphism is using cryptography only for the transformation of data, not to keep the true secret. As the decryption key is stored within every instance of the virus, a virus analyst can always find the contents of the main virus body, simply by the decryption directly, using the stored key. Some viruses use encryption in a more powerful way they encrypt parts of themselves and not including the decryption key within the virus. With only the virus itself and not the decryption key anyone cannot decrypt the hidden code. To be more effective, such a virus must have the way of eventually finding the key, and recognizing that it has found key![10] An early DOS virus, for instance, searched the file system for a filename having the certain checksum, and when that filename with correct checksum was found it used as a decryption key to decrypt certain piece of its own code, which was later executed. Since the encryption techniques were weak and the checksum algorithm was easy to reverse-engineer, anti-virus designers were able to determine that the filename which used as the key was associated with a particular program, and the encrypted code would alter the program's configuration files to introduce a security defect. This types of encryption can make parts of a virus difficult for analysis; however, this techniques has not had any significant impact on the world's information security warfare.

## 2. Related Work

It does not appeared that a properly designed cryptoviral extortion attack has ever been carried out to date immensely. Ted Bridis, wrote an Associated Press article entitled Hackers Holding Computer Files 'Hostage' dated Tuesday. May 24, 2005[5]. This article describes that researchers at Websense Inc. identified the malware infection in which peoples' files are encrypted and held for ransom by the virus author. Symantec has named that malware as Trojan.Pgpcoder. F-Secure analyzed an Trojan (F-Secure Corporation, Technical Details: Alexey Podrezov, May 27-28, 2005) and they referred it as Gpcode. The analysis by F-Secure indicates that this Trojan uses a breakable encryption method. They state that F-Secure Anti-Virus detects that Trojan and repairs the files that it encodes. This is in line with the article, that states that the victim's files were repaired without paying any ransom. Ted Bridis referred this as the "latest threat to computer users" and that it was "unusual extortion plot," overlooking the previous malware that attempted this and the discovered that asymmetric cryptography is needed for carrying the attack correctly [11].

An article also reported that a malware extortion attack has occurred in Europe[6] [12]. The description of that attack is vague. It was not mentioned of public keys, nor asymmetric encryption, nor hybrid encryption. Prior to Trojan.Pgpcoder there were small number of malware attacks that encrypt the host data that were reported by researchers. They all based on symmetric cryptography and hence they were not cryptoviruses/cryptotrojans. Their encryption techniques were therefore reversible and antivirus fixes the problem and decrypted the data they encrypted.[13][14]. Many reports and on-line discussions confuse secure cryptoviral extortion that utilizes asymmetric cryptography with attacks that depend on symmetric encryption alone[03][11][14].

The set of attacks that presented, involve the unique use of strong cryptographic tools in conjunction with computer virus and Trojan horse technology. They also demonstrate how cryptography can allow an adversarial virus writer to gain explicit control over the data that his or her virus has access to, however from an information theoretic point of view[1] this is impossible. The idea is later extended to allow a distributed virus to gain itself explicit access control over the information on infected machines, provided that it is not detected early enough and vigorously destroyed. This shows that viruses can be used as extortion tool, potential criminal activity, and as ammunitions in the context of information warfare. In general, we define cryptovirology as the study of the applications of cryptography for malicious software. It is said that cryptography helps prevent viral attacks and to try to covert a virus's structure, yet formal cryptographic approaches have never before been used successfully as a weapon in the viral attacks[10] . Preventive measures are described in response to those attacks [15]. They are a step toward the direction to help prevent and recover from such attacks. It is fact that the public availability of cryptographic tools without proper access control over them , can put the data on a computer system at serious risk.

## 3. Background

**Cryptoraphy :** "Cryptography" is a benediction to information processing and communications, because it allows people to store information securely and conduct private communications over long distances. There are mainly two aspects in cryptography algorithms and the key used[02]. The important encryption techniques are symmetric , asymmetric, hash function algorithms. Symmetric-key cryptography refers to encryption methods in which both the sender and receiver share the same key of decryption. In asymmetric key cryptography , the encryption key is the public and decryption key is the private key or secret key individually.

**Cryptovirology** : Cryptovirology is the study of the applications of cryptography for implementation of malicious software. It is about, how modern cryptographic paradigms and tools can be used to strengthen, improve, and develop new malicious software attacks. Cryptovirology attacks have been categorized to: give enhanced privacy to the malware and be more robust against reverse-engineering, secondly give the attacker enhanced anonymity while communicating with deployed malware.

**Kleptography:** "Kleptography" is the study of stealing the information more securely. A kleptographic attack is an attack in which a malware designer deploys an asymmetric backdoor. In a this attack, there is an explicit distinction between confidentiality of the messages and awareness of the attack is taking place. A secure kleptographic attack is truly undetectable as long as the cryptosystem is a black-box. A kleptographic attack is an asymmetric backdoor attack that can only used by the designer that carries out the attack. If reverse engineering detects the key generation code then he or she will get known that a kleptographic attack is underway [02].

**Cryptanalysis:** Cryptography is the science of making the secret unintelligible and cryptanalysis is the science of retrieving the secret from that unintelligible data for the further use [03].

## 4. Attacking Methodology of A Cryptovirological Attack

### I. Cryptoviral Extortion [1]:

Cryptoviral extortion is mechanism for the encrypted viruses which uses public key cryptography, in a denial of resources(DoR) attack which can be introduced by the cryptovirus. It is a three-round protocol which is carried out by an attacker against the victim. The attack is generally carried out via a cryptovirus that uses a hybrid cryptosystem to encrypt data while deleting or overwriting the original data in the infecting process. The protocol is as : An asymmetric key pair is generated by the virus designer on a smartcard and the public key is placed in the virus. The private key is especially designated as "non-exportable" so that even the virus author cannot obtain it's bit representation and the private key is generated, stored, and used on the smartcard. Ideally, the smartcard implements two-factor security. Also, the card will ideally be immune against differential power analysis, timing attacks, etc. to prevent the virus author from ever learning those bits of the private key. The virus author then deploys the cryptovirus. After that the virus activates tens or even hundreds of thousands of computer systems. The remainder of this description will cover the protocol for just single such machine.

When the virus get activated, it uses a true random bit generator (TRBG) to generate a symmetric key and initialization vector uniformly at random way. It is essential that the TRBG should produce truly random bits to prevent the symmetric key and the initialization vector from being guessed by the analyst or otherwise will get determined by the victim at a later date. The virus then encrypts host system data with this random symmetric key and the initialization vector. The virus then concatenates the initialization vector with the symmetric key and later  encrypts the resulting string by using the public key of the virus author. The encrypted plaintext is then held for ransom. The virus notifies its victim that the attack has occurred and states that the asymmetric cipher-text will be needed to restore the original data. If the victim ready to pay the ransom and transmitting the asymmetric cipher-text to the virus author then the virus author decrypts the cipher-text using the private key that only the virus author has. The virus author sends that symmetric key and corresponding initialization vector to the victim. These are then used to restore the data that was held ransom. The attack will be is ineffective if the data can be recovered from backups of victim's computer. Antivirus practitioners cannot retrieve the private decryption key by analyzing the virus and from that only the public key will be found. But, this is unacceptable for 2 reasons, the file could be large of size and therefore make the transmission difficult. One question that is generally asked in regards to cryptoviral extortion is: How could an extortionist ever expect to receive the ransom payment? Truly anonymous e-cash could provide a safe way of payment for ransom. This leads to an important consideration for organizations that rely immensely on information technology.

## II. The Secret Sharing Virus [1][16]
This section shows how to implement a virus that is a very close approximation to the highly servile virus. Whereas in the above attack the virus author managed the keys and owned the private key and the virus itself will manage its private key. Since a virus holding a public key and managing its private key can be get analyzed by antivirus analyst and could lose its power. However, this can be accomplished by changing our notion of a system S to be a network of computers, and to regard the host as being the part of entire network. We utilize the distributed environment to hide the key in that virus copies themselves previously. This can be describe in some detail. It is shown that how Public Key Cryptography can be used in a virus to encrypt information in such way that the user cannot retrieve it. In order to be able to decrypt held data to get original, the private key must be storied somewhere, since otherwise held or encrypted data cannot be decrypted. We cannot store the complete private key at one node in the network, as this would give the user of that node the entire private key. By considering an entire network as single host we can effectively divide and conquer the power of the user, since we now have many different users from the network who do not have access to each others data. The secret sharing virus takes advantage of this by sharing its private key among m different nodes, where m > 1. The virus therefore misuse the access controls that users place on themselves to keep its private key secret. The idea is that a virus will then spread itself around the network, and may start acting autonomously or wait for outside control from the author to act as its agent. The local users of the network may wipe out parts of the virus by using back-ups they have, but then the total network might be damaged as we need the entire virus pieces to recover the original data. Then it is useful for the virus to immediately notify all local machines that if they get rid themselves of the virus they may cause global problems(for the whole network) and ask them to consult with the network's administrator.

## III. Deniable Password Snatching
In the DPS attack, the attacker first seeks to install a cryptotrojan into a target computer. Already it seems possible that the attacker is at the high risk of getting caught and most probably if he has installs the cryptotrojan manually. The attack is generally carried out by using a custom cryptovirus designed by the attacker [17]. The attacker(virus author) distributes the virus preferably using the passive virus distribution channel. Once the virus get installed a Trojan horse that it carries with it activated. The purpose of this whole is to allow the attacker to indirectly run code of his own Trojan without being blamed for installing it.

## 5. Attack Scenarios:

**Cryptoviral Extortion**

Scenario 1 is a cryptoviral extortion protocol performed by encryption of the victim's data as ransom. Scenario 2 is same as to Scenario 1 except for that the virus writer also demands the victim's encrypted text for decryption along with the ransom. Scenario 3 explains a secret sharing cryptovirus. The attack works on network with infected hosts. In this attack asymmetric private key is divided into parts and shared among all infected network hosts. Scenario 4 states the involvement of a cryptovirus in a Deniable Password Snatching (DPS) attack . [2][15][16][18][23].

▫**Scenario 1:**

The cryptoviral extortion protocol is performed by using a cryptovirus. It is the three round protocol between virus author and the victim. The attack based on hybrid encryption scheme that works by encrypting the original data as well as overwriting or deleting that data on the victim's computer. A pair of asymmetric key is generated by the virus author and merge the public key into virus itself. The corresponding private key is only with the virus author.

1.First, author introduces his virus on a public communication medium such as the Internet. More and more hosts get infected by the virus if no safeguards are in place in the network. Once on the victim's machine, the cryptovirus sets into action. It generates a random symmetric key and Initialization Vector. The user's data on the secondary storage device is then encrypted using this key and later get chained using a chaining mode as CBC. The actual information is then may be deleted or overwritten. The Initialization Vector is appended with the symmetric key and encrypted using the virus author's public key. The encrypted plaintext is then held for ransom. After infection the encrypted plaintext and anonymous ways to contact the virus writer are displayed on the victim's screen.

2. If the victim agrees on the condition to pay demanded ransom, he should transmits the encrypted pair to the virus author. The virus writer then decrypts the pair by using the corresponding private key and sends back the pair to the victim.

3. Using this pair, the victim is now able to decrypt the held information on secondary storage medium. The attack can be ineffective if the data is recovered from backups. It is however common thing that many organizations have a daily or weekly based backup policy. In those cases the cryptoviral attacks can still valid but for a shorter period of time. The effectiveness of the attack also rely on the sensitivity of the information and the amount of working hours required to duplicate the result. In practice a compact and fast code encryption algorithm such as the Tiny Encryption Algorithm (TEA) can be used for symmetric key encryption. A hybrid encryption scheme is used in this case so as to increase the chances of the victim cooperating to pay the ransom. In the above scheme, the virus writer never comes to know the data content that are held at ransom. In this a model of a scheme the thief profits, though he doesn't take anything. Another interesting question is how the virus writer will receive cash payment anonymously.

▫**Scenario2:**
The cryptovirus can be used as a secret sharing virus. The idea works in the following way: We have a local network containing n hosts. The mode of attack is same as the previous scenarios except for that RSA private key RSApri(RSA private key)is now split among k or more nodes, k<n and RSApri no longer resides with the virus writer. The advantage is that the virus author no longer needs to handle the RSA private keys and the victim need not send anything other than the ransom to the virus author. The secret sharing scheme takes advantage of the fact that the RSA private key will split among k or more hosts on the network making it hard for the system administrator or the virus analyst to find them and recover them back[14].

▫**Scenario3:**
We explore the role of a cryptovirus in a Deniable Password Snatching (DPS) which is commonly used in espionage. An attack is said to be DPS if the attacker is can recover passwords from the system in the such a manner that the attacker is untraceable. I.e. even if the malware get discovered, the virus author cannot be proven guilty. The another condition for DPS is that the stolen login/password pairs are only accessible by the attacker and no one else including administrator of the infected host. A DPS attack can performed in following way:

First, the attacker needs to put a Crypto Trojan into the targeted host. If he directly goes and tries to inserts the cryptotrojan, he has a greater chance of being caught in this act. Moreover evidence can be collected pointing towards the attack should worsen and attackers chance of being convicted for the crime. In other way, the attacker chooses a passive channel such as IBB(Internet Bulletin Boards) or the infected CD's or diskettes that are swapped without known to victim. This is ensures that the cryptotrojan is not able to traced back to its author and cannot be held responsible for attack. It is interesting how the cryptotrojan makes use of the El-Gamal Cryptosystem to make those attack work. [8].

▫**Scenario4**
The cryptovirus can be act as a secret sharing virus. This idea works in the following way: There is a local network with *n* hosts. The mode of attack is same as to the previous scenarios except for that RSA private key RSApri is split among *k* or more hosts, *k<n* and RSApri no longer resides with the virus writer. The advantage of this is, the virus writer no longer necessary to handle the RSA private keys and also the victim need not send anything other than the ransom to the attacker. This scheme takes advantage of the fact that the RSA private key will be divided across *k* or more nodes on the network making more difficult for the system administrator find them and regain the original data[22].

## 6. Latest Events:

▪**GPCode[03][4][19]:**
This Trojan modifies data on the victim system so that the victim can not use the data or it prevents the computer from functioning correctly. Once the data has been encrypted, the user will receive demand for ransom. The ransom demand states to send the money to the attacker; on the receipt of this, the virus writer will send a program to the victim to restore the original data or restore his computer's performance. gpcode is one of the classic cryptovirus. A virus that informs victim to pay a ransom is the virus nicknamed Tro_Ransom.A after hostage [5]. This virus asks the victim to send $10.99 to a given account by Western Union account. Another classic cryptovirus is Virus.Win32.Gpcode.ag. It partially uses the version of 660-bit RSA(RSA private key) and encrypts files in many different extensions and overwrite or removes the original one . It tells victim to email a given mail ID if the he desires the decryptor of data to regain the original data. If attacker contacted by email, the victim will be asked to pay a certain amount of money as ransom in return for the decryptor of data.

**How Gpcode spreads by E-mail:**

**Example:**

**E-mail 1:**
Hello!

We are writing to you regarding the resume you have submitted on our site job.ru. I have a vacancy in the ADC Marketing LTD (UK) that is suitable for you. Company is opening a new branch office in Moscow and I am searching for appropriate candidates for the positions in the company. Soon I will ask you for an walk-in interview on the mutually convenient time. If you are interested in this offer, please fill the form attached related to compensation and email it to me.

Sincerely,

Viktor Pavlov, HR manager

The file attached to the above email was MS word .doc file called as anketa.doc attached. (Anketa means the Russian for application form)

▪ File contains a malicious program called Trojan.

▪On opening the file, a malicious macro installs another Trojan, TrojanDownloader.Win32.Small.crb on the victim's machine.

▪ This Trojan then downloads Gpcode from [skip].msk.ru/services.txt and installs it to the

   victim machine.

▪ It scans all accessible directories and encrypts files with certain extensions such as .txt,

   .xls, .rar, .doc, .html, .pdf etc.

▪ also encrypts mail client databases.

▪ Gpcode and the other trojans self destruct.

▪ Leave behind a file in each directory which has an Encrypted file with README.txt,

   Some files are coded by RSA method.

To buy decoder mail: k47674@mail.ru  with subject: REPLY

▪ WaterLilles.jpg - original file

▪ WaterLilles.jpg. - CRYPT — encrypted file

**Email 2:**
'Next, you should deposit $100 to Liberty Reserve account U6890784 or E-Gold account 5431725 (www.e-gold.com). To   buy E-currency you may use it in the exchange service. Specify your e-mail in the transfer description field.

After receiving your payment, we will send decryptor to you through your e-mail. To check our guarantee you can send us one any encrypted file (with cipher key which is specified in any !_READ_ME_!.txt file are in the directories with the encrypted files) for the demo. We will decrypt it and send to you some part of originally that decrypted file.

Best Regards,

   Daniel Robertson

▪**Conficker[03][15][16][20][24]:**
Conficker is also nicknamed as Downup, Downadup and Kido, is a computer worm targeted to the Microsoft Windows operating systems that was detected in November 2008 for first time. It make use of the flaws in Windows software and make Dictionary attack on administrator passwords to co-opt system and link them into a virtual computer that can be commanded remotely by attacker. Conficker has since spread very rapidly into what is now known as the largest computer worm infection since the 2003 SQL Slammer, with more than seven million government, business and home computers in over two hundred countries now in its control[14][4]. The worm has been usually difficult to detect because of its combined use of many advanced malware tools and techniques. As shown in figure1 it compute a 512-bit hash M of the windows binary. That Binary is then encrypted using the algorithm symmetric stream cipher RC4 which is having password M. Lastly the Digital signature is calculated using the RSA(RSA private key) encryption scheme. Sig N is a public modulus that embedded in all the Conficker client's binaries. then Sig is appended to the encrypted binary file, and either they may be pushed it to all the infected Conficker clients.

Figure1:Binary Download and Validation of Conficker

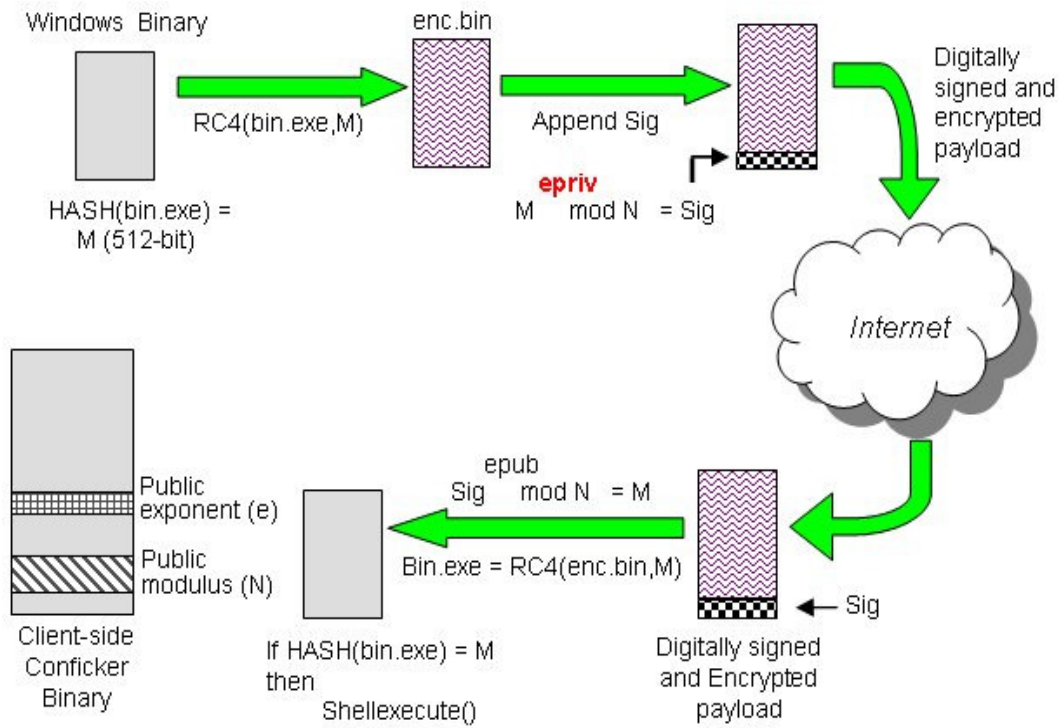

**Timeline[14]:**

▪ Win32/Conficker.A / November 21, 2008.

▪ Win32/Conficker.B / December 29, 2008.

▪ Win32/Conficker.C / February 20, 2009.

▪ Win32/Conficker.D / March 4, 2009.

▪ Win32/Conficker.E / April 8, 2009.

**Initial Infection**
Variants A, B, C and E mainly exploit a vulnerability in the Server Service on the Windows computers. The worm runs an HTTP server on a port between 1024 and 10000 in the source machine. The target shell-code connects back to this HTTP server to download a new copy of the worm in the form of DLL file, which then attached to a running service, as shown.

**Domain Name Generation**
 Conficker implements its own random number generator(Might be TRBG). Alternately chooses between its own functions generate_random() and the system's default rand() function. The domain suffix is chosen randomly between .com, .net, .org, .info, and .biz is then appended to domain name. Conficker B encompasses additional suffixes as (.ws, .cn, .cc).

## 7. Counter Measures
There are several measures that can be taken to immensely reduce the risk of being get attacked or infected by a cryptovirus. Many of the attacks described in here can be avoided with existing antiviral mechanisms and tools, since cryptoviruses propagate in the similarly as traditional viruses does. The first step toward this is, implementing mechanisms and schemes to detect viruses before infection or may be immediately following system infiltration[1][15][21].

1. The direct memory monitoring is needed to catch self polymorphic encrypting, cryproviruses.

2. Only passwords are weak to authenticate a host. Two level authentication, first with the biometric entity such as the users fingerprint, iris scan and second using the password .

3. Use of up-to-date anti-virus mechanism, as the crypto viruses spread similar to normal viruses does.

4. Provide an access control mechanism to cryptographic tools and API's of the system. This will helps virus analyst or the system administrator to identify suspicious cryptographic usage of the system component.

5. There are strong cryptographic techniques and good pseudo random number generators like True Random bit Generator(TRBG) are available, it makes the virus simpler and simpler to implement and more faster to execute if the code is optimized. Use of powerful cryptographic tools in the operating systems might increase system security but makes the system vulnerable since viruses can call these operation system routines for malicious purpose. Hence it is necessary to monitor the processes that invoking the cryptographic routines and make attempt to prevent and lock processes that do not have enough access privileges to call the cryptographic routines.

6. Use of intrusion detection System and firewall to protect host in the network system and the stand alone systems. Use patches as they are made available to make system more secure.

## 8. Role of Cryptovirology In Information Warfare:
Battles of the future information warfare will be decided by the countries which have the leading edge in cryptovirological technologies and its countermeasures. This may be used to create panic by using methods such as rising a false nuclear alarm, may be used to block and encrypt military databases of the enemy nations. Also Can be used to bring down communication in networks of enemies nation by causing Denial-of-Service (DoS) attacks in large scale [02][10].

## 9. Why To Learn Cryptovirology ?

A cryptovirological attacker attacks a computer or the network of computer in the same wayas a cryptanalyst attacks a cryptosystem. Should we stop trying to analyze cryptosystems and hope that they will be secure automatically? Of course not , by the same token we should not stop trying to analyzing what attackers might do once they break into systems. The justification of doing research in cryptovirology can be derive from the proverbial phrase, "It takes a thief to catch a thief". The notion is that thieves are the experts when someone comes to thieving, and they would know the best regarding how to catch other thieves and what are the ways he might use. Cryptovirology is a proactive anticipation of the opponent's or attackers next move and suggests that certain safeguards against it[02].

## 10.An Implementation And Performance Of The Cryptovirus

To successfully implement a cryptovirus, a through the various cryptographic approaches such as random number generators(TRBG), proper recommended cipher text chaining modes such as CRC etc. are necessary. Wrong choices can lead to poor cryptographic attack and increase chances of getting caught. So, use of previously existing cryptographic routines would seems to be ideal such as Microsoft's Cryptographic API (CAPI) [11][22]. It has been shown that using just eight different calls  API calls, a cryptovirus can satisfy all its encryption requirements.

**Pre-computation**: Two pre-computations were performed to create the designated cryptovirus. The only other form of virus that we are aware of, requires pre-computation to decompress itself during runtime. Such viruses save on secondary storage and are therefore slightly traceable. The first pre-computation performed was for the generation of a public and the private key pair. The key modulus used were different combination 512 long bit stream [1].

**Algorithm:** There is a special algorithm is used to perform the modulo operation for RSA(for primary key) encryption. This method is a based on the division algorithm of the repeated subtraction method [14].

**Operation of the Virus:** It can be exists in any of three states at any given time may be in a program, in the operating system file, or in a patch to an operating system routine such as API. When an infected program on the system is run, the virus gets control prior to host program and checks that if the system file is already infected or not. If it is uninfected, the system file gets infected and control is then returned to the system by the virus. Next time when the system is rebooted, the virus copies itself into RAM and make modifications in the trap dispatch table so that the table invokes the resident copy of virus whenever that program is run. Once a program is run, the virus that resides in the patch (according to dispatch table) will check if the program is already infected or not. If it is uninfected, the virus will infect it and so on. This operation destroys or overwrites the original file, then the virus program overwrites the RSA plaintext key in RAM, and creates the new file entitled as 'VIRUS DAMAGE' in system folder. This file contains about on how to contact the attacker and the RSA ciphertext  would be used  later for decryption [1].

**Performance:** The following table is a summary about the performance of the cryptovirus related to the time in the infection [1].

Table1.Running Time

| System boot(normally) | < | 16.7 msec |
|---|---|---|
| Generation of 384 random bits | = | 6.4 sec |
| Infect a System file | ≈ | 4 sec |

| RSA Encryption | = | 66.7 msec |
|---|---|---|
| System boot(w/attack) | = | 11.92 sec |
| Infect a program | ≈ | 1 sec |
| True Encr. Algorithm Rate(1 round) | = | 47k bytes/sec |
| TEA Rate (3 rounds) | = | 15.7k bytes/sec |

Note that about ten minutes of CPU time was spent on the above pre-computations. The approximate running time are given because they can vary program to program. Factors such as disk response time can make the variations. The critical file and desired files used for this benchmark were each 30 kb approximately in the length. There are no disk writes needed in the system boot phase of the virus, however disk writes are needed in infection operation. And because of that the system boot phase takes significant less time in second case. The random number generation takes up 53.7 percent of the total attack time.

Table2.Virus size

| Code size | bytes | Source code language |
|---|---|---|
| The main Attack routines | 432 | ANCI C |
| Global data | 560 | N/A |
| TEA encryption routine | 88 | ASM |
| Modified GNU MP lib | 4,372 | ANSI C |
| Misc attack code | 804 | ANCI C |
| Main virus routine | 614 | ASM |
| Entire attack routine | 6,372 | ANCI C/ASM |
| True rand() size | 124 | ASM |

It can be inferred from above table that the attacking routine could be made much small if they were written in machine language. The outcomes of our research (by Adam Young and Moti Yung) was that we found that it is possible to write code for RSA, true_rand(), and True Encryption Algorithm, such that code s not exceed 7kb in size. Optimizing the size of code was a challenging since many viruses are considerably small in size. Optimizations allowed to omit exponentiation, multi-precision and a division routines. This optimizations were used in areas like smart card technology [1].

## 11. Future Scope:
Consideration of what the nature of viral attacks might will be in the future is important to successfully protecting against them. The understanding of how cryptography engaged in the cryptovirology and necessary management techniques can definitively checkmate future attacks. Traditionally, cryptography is for the defensive in purposes, and provides privacy in communication of data, authentication in computer systems, and to provide high level of security to users. The analysis of potential threats and attacks that misuse cryptography can be immensely threatening when get combined with malicious software like viruses, Trojan horses. The public-key cryptography is necessary in the attacks that are demonstrated by the researchers(Adam Young and Moti Yung)[02][03]. Also the countermeasures and mechanisms can be suggested to defend or prior prevention of such attacks. These attacks have shows how cryptographic techniques and tools should be audited and managed in general computing environments, and demonstrates that access to the cryptographic tools should be well controlled such as system's built-in cryptographic routine(API routines). Experimental viruses demonstrates that cryptographic packages can packed into very small size, which might have independent applications in the future information communication (e.g., cryptographic module designed for small mobile devices).

## 12. Conclusions

It is demonstrated that how Cryptography can be used to built the malicious software programs that can be used in extortion-based attacks on computer system. Public-key cryptography is necessary for the attacker to take advantage over the owner of infected system. Also presented an demonstrative cryptoviruses which could accomplishes this very easily. And a set of measures that can be taken to minimize the risks of attack posed by the cryptovirology. In short, the attack based on a cryptotrojan or cryptovirust, if made then it is not possible to linked it to its author. The cryptotrojan take the control over internal resources to steal passwords and login information, and make use of the strengths of probabilistic public key cryptography to maintain exclusivity of the stolen (or encrypted) data with respect to the virus author. The cryptotrojan is elusive in nature as it prevents the system administrator from being able to prove that anything has been stolen when the cryptotrojan is discovered due to its statelessness nature. Though the attackers actions on the victim's computer are monitored, it still not reasonable to prove that he or she is the attacker; it is called an "originator concealing" attack. Evidence should be collected from the "attackers computer" only.

We conclude can from all these attacks that auditing and logging tools may inadequate for the law enforcement purposes against the attacker. From the specific attack on passwords it can be further strengthen the view that there is an inherent weakness in expository authentication techniques and need to more secure systems which may use hardware tokens, one-time passwords scheme, and other identification protocols. Finally, Cryptographic techniques and tools can be used to create a new class of viruses: Cryptoviruses in which the virus author need not be aware of underlying cryptography and use of available functions(like API). It will be better to make attacks publicly known rather to wait for attacks to occur.

## Acknowledgement

I am thankful to Management of Vishwakarma Institute of Information Technology, Pune and all my faculty members of Computer Engineering Department for their encouragement and whole hearted cooperation. I would like to thank Prof. Y. V. Dongare for his guidance and valuable suggestions time to time. I am lucky to have god like Parents, they provided me very excellent moral support. I would like to express profuse thanks to all my colleagues for their support during this work.

International Journal of Network Security & Its Applications (IJNSA), Vol.3, No.4, July 2011

**Author:**

**Shivale S. A.** (shivalesaurabh@yahoo.com), is pursuing BE (Computer Engineering) from Vishwakarma Institute Of Information Technology, Pune-48, India. Presently in third year and has completed diploma in Computer Technology in Government Polytechnic, Nashik. His area of interest includes Network & Information Security and development technologies.

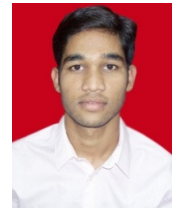